\documentclass[twocolumn,prl,aps,byrevtex]{revtex4}
\usepackage{graphics}
\usepackage{amsmath,amssymb}
\newcommand{\GI}{GI }

\newcommand{\x}{{\bf x}}

\newcommand{\xp}{{\bf x}'}

\begin{document}
\noindent \textbf{Comment on ``Entangled-Photon Imaging of a Pure
  Phase Object''} 

Recently Abouraddy \textit{et al.} showed experimentally the ghost
interference (GI) of a pure phase object using entangled photons
\cite{abouraddy:2004etal}. Our theory showed that the same result can be
obtained using beams created by dividing a thermal-like speckle beam
on a beam splitter (BS) \cite{thermal}. This realizes \textit{coherent
  ghost imaging using spatially incoherent classical beams}: the
interference pattern of an object (phase or amplitude) appears in the
correlation between the two arms, while no interference appears in a
direct detection of the arm containing the object [see Eq.~(16)
of \cite{thermal}(b)]. 
Our recent experiment confirmed this \cite{ferri:2004}, but
%We validated these results experimentally \cite{ferri:2004}.  
%However,
the introduction of \cite{abouraddy:2004etal} misinterprets that
experiment as to use spatially coherent (rather than incoherent)
beams. Thus, they claim, \GI of a phase object may occur using
thermal beams only because the beams have second order spatial
coherence, whereas the entangled photons in \cite{abouraddy:2004etal}
have no second order spatial coherence.  This Comment shows that such
an interpretation is not correct.
\par
\cite{abouraddy:2004etal} claims that in our experiment the diaphragm
``provides a large coherence area. This configuration thus effectively
endows the thermal light with second order, as well as fourth order,
coherence''. On the contrary, as stated in \cite{ferri:2004} the
diaphragm (diameter $D=3$ mm) is much larger than the speckle size there ($\Delta
x\simeq 25~\mu$m), thus selecting a large number of speckles
[estimated by $N_{\rm sp}\equiv (D/\Delta x)^2$].
 Dividing this
\textit{spatially incoherent} speckle beam on a BS creates two copies,
each on its own incoherent. However, since they are
identical speckle-by-speckle they have a high \textit{mutual spatial
  correlation} preserved on propagation \cite{thermal}.
%: this is the scheme's key feature.  
Really crucial  is  that the speckles at the object plane ($\Delta
x_n\simeq34~\mu$m) are much smaller than the object ($690$ $\mu$m ), which provides incoherent illumination of the object. Only in this case GI can be performed with high resolution\cite{thermal}. 
%In
%fact, the speckle size governs the resolution of GI: the more incoherent the
%input beam the better the resolution \cite{thermal}.
% Instead, the Hanbury-Brown--Twiss interferometer
% \cite{hanburybrown:1956} for determining the stellar diameter relies
% on coherence gained by propagation.

Entangled ghost imaging has an \textit{analogy} (rather than a
duality) in thermal ghost imaging \cite{thermal},
explaining why \GI of a phase object may occur with classical
incoherent beams. In each case the intensity fluctuation correlation
$G=\langle I_1
I_2\rangle-\langle I_1 \rangle\langle I_2\rangle$ between arm 1 and 2 %namely
obeys \cite{thermal}
\begin{eqnarray}
  \label{eq:pdc}
G_{\rm ent}(\x_1, \x_2)  \propto
\nonumber\\\times\left| \iint d \xp_1
d \xp_2  h_1 (\x_1, \xp_1) h_2 (\x_2, \xp_2) \Gamma_{\rm
  ent}(\xp_1,\xp_2) 
\right|^2,\\
  \label{eq:th}
  G_{\rm cl}  (\x_1, \x_2) \propto
\nonumber\\
\times
\left
| \iint d \xp_1 d \xp_2  h_1^* (\x_1, \xp_1) h_2 (\x_2, \xp_2)
\Gamma_{\rm 
  cl}(\xp_1,\xp_2)
\right
|^2.
\end{eqnarray}
$\Gamma$ is the field correlation at the object plane,
% $r$ and $t$ the BS reflection and transmission coefficients. 
$h_1$ and $h_2$ 
are the impulse response functions of the two arms.
%, one of which contains the object. 
The two cases are analogous: (a) Both are \textit{coherent imaging
  systems}, implying both may show GI of a phase object. (b)
Both perform similarly if $\Gamma_{\rm cl}$ and $\Gamma_{\rm ent}$
have similar properties. A numerical  example of GI of  a pure phase object with 
thermal-like beams is shown in \cite{thermal}(a).
%As \cite{thermal} showed, they differ in the
%visibility (ideally unity in the entangled case): while below
%$\frac{1}{2}$ in the classical case, good visibility can be achieved
%to efficiently retrieve information \cite{thermal,ferri:2004}; in
%particular Fig.~2 of \cite{thermal}(a) shows numerically \GI of a pure
%phase object with classical beams.  Thus no entanglement is required
%for this.
%This is not difficult to achieve, as already
%demonstrated in \cite{thermal,ferri:2004}.
\begin{figure}[t]
{\scalebox{.38}{\includegraphics*{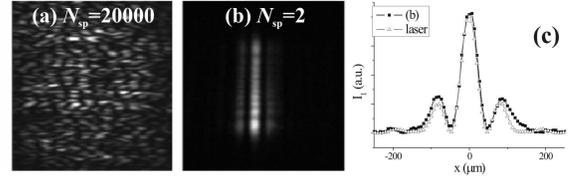}}} 
\caption{Single shot intensity distribution $I_1$ in the far field of the object. Similar setup as in \cite{ferri:2004}, but in (b),(c) the spatial coherence of the light illuminating the object is increased  by reducing the source size.
  (a) Source size $D_0\simeq 10$ mm, speckle size at the diaphragm $\Delta
  x\simeq 21~\mu$m,  $N_{\rm sp}\simeq 2\cdot 10^4$.
  (b) A small pinhole (0.1 mm) right after the ground
  glass gives $D_0\simeq 0.1$ mm, so that $\Delta x\simeq 2.1$ mm, 
  $N_{\rm sp}\simeq 2$. (c) $x$-cut of (b) compared with the
  diffraction pattern with laser illumination.  
% but to
%   increase the power a Nd:YAG laser ($\lambda=532$ nm) was used and
%   the turbid medium was absent.
}
\label{fig} 
\end{figure}
%No \GI will occur if the diaphragm selects a single speckle

Had the diaphragm selected a single speckle then
no \GI would have occured: %, because
$\Gamma_{\rm cl}$ can then be considered constant in space, and the
two integrals in~(\ref{eq:th}) factorize to a product of two ordinary
imaging schemes, showing the diffraction pattern of the object only in arm 1. This is no longer \GI since no diffraction
pattern of the object is observed by scanning the detector position in arm 2. 
Our experiment indeed used incoherent beams, as shown in Figs.~3(a)
and 4 of \cite{ferri:2004}. Figure~\ref{fig} shows the transition from
the incoherent case of \cite{ferri:2004} to the coherent case. In
Fig.~\ref{fig}(a) the setup is the same as in \cite{ferri:2004},
except that a Nd:YAG laser %($\lambda=532$ nm) 
is used and the turbid medium is removed to increase the power. The
diaphragm transmits many speckles ($N_{\rm sp}\simeq 2\cdot 10^4$, $\Delta x\simeq 21~\mu$m ). Because of light incoherence no diffraction pattern appears in the far field of the
object beam, neither in the single-shot intensity distribution  $I_1$ [as in Fig.~3(a) of \cite{ferri:2004}], nor in its average over frames. Conversely, 
in Fig.~\ref{fig}(b),(c) we increase the speckle size by reducing the source size (Van Cittert-Zernike theorem), to the point where the diaphragm roughly selects a single speckle ($N_{\rm sp}\simeq 2$). The beam becomes spatially coherent and the object diffraction pattern clearly shows up in $I_1$.
This latter case is completely different
from what we observed in \cite{ferri:2004}.

% (as a result of the factorization). 
%: the conditions for \GI are no longer there.
% that the transition from coherent to 
% incoherent imaging by increasing the spatial coherence of the input
% speckle beam (by reducing the size of the pinhole). The large pinhole
% (as in \cite{ferri:2004}) gives incoherent beams but coherent imaging: a
% ghost diffraction is observed from the correlation, while no
% diffraction pattern is recorded in the intensity of the object arm
% $\langle I_1(\x_1)\rangle$. Decreasing the pinhole size gives more and
% more coherent beams, while the imaging coherence decreases: therefore
% the ghost diffraction disappears while a diffraction pattern appears
% in $\langle I_1(\x_1)\rangle$.
                          
This work was supported by COFIN of MIUR, INTAS 2001-2097, and the
Carlsberg Foundation.\\

\noindent M. Bache, A. Gatti, E. Brambilla, D. Magatti, F. Ferri, and
L.A. Lugiato

{\small INFM, Dipartimento di Fisica e Matematica, Universit{\`a} dell'Insubria, Como, Italy}

\noindent December 20, 2004

%\bibliographystyle{C:/texmf/tex/latex/revtex4/apsrev}
%\bibliographystyle{d:/LocalTexMf/miktex/prsty}
%\bibliography{d:/Projects/Bibtex/literature}

\end{document}